\documentclass[pra]{revtex4}
\usepackage{amsmath,bm,amssymb}
\usepackage{color}
\usepackage{braket}
\usepackage{graphicx}
\makeatletter

\newcommand \be {\begin{equation}}
\newcommand \ee {\end{equation}}

\newcommand{\I}{\mathrm{i}}
\def\tr{\mathrm{tr}}
\def\e{\mathrm{e}}
\def\thetaone{\theta_1}
\def\thetatwo{\theta_2}
\def\thetai{\theta_{\, i}}
\def\ka{\kappa_\mathrm{a}}
\def\kb{\kappa_\mathrm{b}}
\def\Gspp{G_\mathrm{S}^p}
\def\Gspip{G_\mathrm{S}^\pi}
\def\GGrpp{\tilde{G}_\mathrm{R}^p}
\def\GGrpip{\tilde{G}_\mathrm{R}^\pi}
\def\Gspt{G_\mathrm{S}^{ p \, \mathrm{thermal}}}
\def\Gspit{G_\mathrm{S}^{ \pi \, \mathrm{thermal}}}
\def\Grpt{G_\mathrm{R}^{ p \, \mathrm{thermal}}}
\def\Grpit{G_\mathrm{R}^{ \pi \, \mathrm{thermal}}}

\def\GGrpip{\tilde{G}_\mathrm{R}^\pi}

\begin{document}

\title{Uncertainty relation for estimating the position of an electron in a uniform magnetic field from quantum estimation theory}
\author{Shin Funada}
\author{Jun Suzuki}
\affiliation{%
Graduate School of Informatics and Engineering, The University of Electro-Communications, 
1-5-1 Chofugaoka, Chofu-shi, Tokyo, 182-8585 Japan
}%
\date{\today}

\begin{abstract}
We investigate the uncertainty relation for estimating the position of one electron in a uniform magnetic field in the framework of the quantum estimation theory. 
Two kinds of momenta, canonical one and mechanical one, are used to generate a shift in the position of the electron. 
We first consider pure state models whose wave function is in the ground state with zero angular momentum. 
The model generated by the two-commuting canonical momenta becomes the quasi-classical model, in which the symmetric logarithmic derivative quantum Cram\'er-Rao bound is achievable. 
The model generated by the two non-commuting mechanical momenta, on the other hand, turns out to be a Gaussian model, where the generalized right logarithmic derivative quantum Cram\'er-Rao bound is achievable. 
We next consider mixed-state models by taking into account the effects of thermal noise. 
The model with the canonical momenta now becomes genuine quantum mechanical, although its generators commute with each other. 
The derived uncertainty relationship is in general determined by two different quantum Cram\'er-Rao bounds in a non-trivial manner. 
The model with the mechanical momenta is identified with the well-known Gaussian shift model, 
and the uncertainty relation is governed by the right logarithmic derivative Cram\'er-Rao bound. 
\end{abstract}


\maketitle

\section{Introduction}
The uncertainty relation based on the quantum estimation theory was investigated 
by many authors, see for example 
\cite{Helstrom,holevo,braunstein,nagaoka,gibilisco,watanabe,gzjfw2016,kull}.  
It is known that one-parameter unitary model with a pure reference state, the Heisenberg-Robertson type uncertainty relation and 
the uncertainty relation by the parameter estimation have the same form. 
Further, this type of approach is more general than the traditional one, since one can derive the uncertainty relation 
for non-observables. The celebrated energy-time uncertainty relation \cite{mt45} is a well-defined relation for time and energy when treated by the quantum estimation theory.
In literature, many authors discussed similarity between two-different types of uncertainty relations. 
In Ref.~\cite{watanabe}, they showed that the uncertainty relation for a generic full parameter qudit model 
can be different when derived from the quantum parameter estimation theory. 
Usually, when the uncertainty relation is discussed, 
the uncertainty relation of two non-commuting observables is discussed, see for example~\cite{ozawa,ozawa2,wehner}. 

The aim of this paper is to investigate the uncertainty relation between two {\it communing observables} based on the multi-parameter quantum estimation theory \cite{holevo,paris,alb,sidhu,RDD}. 
At first sight, one might expect that there cannot be such a trade-off relation. However, as demonstrated in this paper, 
the quantum estimation theory enables us to derive a non-trivial trade-off relation for estimating the expectation values of two commuting observables.
In the present work, we set up a specific physical model, a model of one electron in a uniform magnetic field 
and investigate the uncertainty relation regarding the position of the electron by the parameter estimation problem 
of two-parameter unitary model.
In this model, the Heisenberg-Robertson type uncertainty relation~\cite{heisenberg,robertson} of the position operators $X,Y$ of an electron only yields the following trivial inequality. 
\be
( \Delta X )( \Delta Y ) \geq \frac{1}{2} | \langle [X, \, Y] \rangle_{\rho} | = 0. 
\ee
This is because two position operators $X$ and $Y$ commute, i.e., $[X, \, Y] = 0$. 
In the relation above, $\Delta X$ denotes the (quantum) standard deviation about $X$ with respect to a state $\rho$, 
which is defined by 
\be
(\Delta X)^2 = \tr \, [ \rho (X - \langle X \rangle_{\rho} )^2 ] = \langle X^2 \rangle_\rho - \langle X \rangle^2_\rho, \label{delta_x_sq}
\ee
with $\langle X \rangle_\rho = \tr \, [ \rho X ]$ the expectation value of $X$. $\Delta Y$ is defined similarly.

In order to derive the uncertainty relation between $X$ and $Y$, 
we need to introduce a parametric model describing the position measurement of the electron.
We use the unitary transformation generated by the canonical momenta 
$p_x$ and $p_y$ with the parameter $\theta = (\thetaone, \, \thetatwo)$.  
The state $\rho^p_\theta$ generated by this transformation from the reference state $\rho_0$, 
which is known in advance, is defined as
\be
\mbox{Model 1:}\quad \rho^p_\theta = \e^{- \I \thetaone p_x} \e^{- \I \thetatwo p_y} \rho_0 \e^{ \I \thetatwo p_y} \e^{ \I \thetaone p_x} .
\label{c_ope11}
\ee
Using the generators $p_x$ and $p_y$, the expectation values of the position operators are 
\begin{align}
\langle X \rangle_\theta &= \langle X \rangle_0 + \thetaone, \label{shift1} \\
\langle Y \rangle_\theta &= \langle Y \rangle_0 + \thetatwo. \label{shift2}
\end{align}
where
$\langle X \rangle_\theta = \tr \, [\rho^p_\theta \, X]$ and $\langle X \rangle_0 = \tr \, [\rho_0 \, X]$.  
We define $\langle Y \rangle_\theta$ and $\langle Y \rangle_0$ similarly. 
From Eqs.~(\ref{shift1},~\ref{shift2}), we see that estimating the parameters $\theta_1$ and $\theta_2$ 
amounts to the measurement of the position $X$ and $Y$. 
It is worth noting that the generators of Model 1 also commute, i.e., $[p_x, \, p_y]=0$.

As the main contribution of this paper, we derive an uncertainty relation, or a trade-off relation between 
the components of mean square error (MSE) matrix by using two different types of the quantum Cram\' er-Rao (C-R) inequalities for Model 1. 
In particular, we find a structure change in the uncertainty relation and derive a condition for this transition analytically.

We can generate another shift model of the position density probability that gives the same relation as Eqs.~(\ref{shift1},~\ref{shift2}).  That is
\be
\mbox{Model 2:}\quad \rho^{\pi}_\theta = \e^{- \I \thetaone \pi_x} \e^{- \I \thetatwo \pi_y} \rho_0 \e^{ \I \thetatwo \pi_y} \e^{ \I \thetaone \pi_x},
\label{c_ope22} 
\ee
where $\vec{\pi} = \vec{p} + e \vec{A}$.   The vector potential for the uniform field $\vec{B}$ is denoted by $\vec{A}$. 
The charge of an electron is $-e \: (e > 0) $. 
We use Model 2 as a reference, since we can map the model to a well-studied Gaussian shift model \cite{holevo}.   
The Hamiltonian $H$ of this system has an equivalent form of the harmonic oscillator with respect to the generators $\pi_x, \, \pi_y$ \cite{johnson}. 
Both Model 1 and Model 2, therefore, make a shift in the position 
of the position probability density which is defined by the product of the wave function and its complex conjugate. 
However, there is a significant difference between these two models. 
The generators of Model 2 do not commute, $[ \pi_x, \,  \pi_y]= -\I eB$ unlike those of Model 1. 
Furthermore, Model 2 $\rho^{\pi}_\theta$ defined by \eqref{c_ope22} turns out to be a displaced Gaussian state when $\rho_0$ is a thermal state. 

The outline and the summary of this paper are as follows. 
In Sec.~\ref{sec:phy_model}, the Hamiltonian of the system is given in terms of the creation and annihilation operators. 
In Sec.~\ref{sec:xyMeasQuest}, we explain how the position measurement of the electron can be 
set up as a two-parameter estimation problem. In Sec.~\ref{sec:URQuest}, we derive the uncertainty relation 
for the MSE matrix for arbitrary two-parameter estimation problem from the quantum C-R inequality.

In Sec.~\ref{sec:Pure_st}, 
we investigate the trade-off relation by estimating the position of electron 
with respect to the reference state which is a pure state.  As the reference state, 
we choose the lowest energy state with zero angular momentum, 
or the lowest Landau level (LLL) \cite{landauQM}. 
The position probability density of the LLL is known to be a Gaussian function 
$\propto \exp[- (x^2 + y^2)/\lambda^2]$ 
where $\lambda = \sqrt{2(eB)^{-1}}$ has the dimension of length.
We obtain the uncertainty relation from the symmetric logarithmic derivative (SLD) C-R inequality for the MSE matrix,
which cannot be less than ${\lambda^2} / {2}$ for Model 1.  
The measurement accuracy is limited by the spread of the position probability density of LLL.  
For Model 2, in the meantime, the generalized right logarithmic derivative (RLD) C-R bound 
\cite{fujiwara2,fujiwara3,fujiwara4} sets the achievable bound for the MSE matrix.  
Thereby, we show that the C-R bound of Model 2 is lower than that of Model 1, 
indicating that Model 2 potentially gives more accurate way of estimating the position of the electron.  

In Sec.~\ref{sec:Mixed_st}, as the other choice of the reference state, we use a thermal state to see effects of noise.  
In this system, an infinite number of the angular momentum eigenstates exist at each energy eigenstate.  
The energy eigenstate of this system is degenerated. (See section Sec.~\ref{sec:degeneracy}.)
To avoid possible problems caused by the degeneracy, we impose a condition that the expectation value of the angular momentum 
$\langle L \rangle_0$ is fixed. 
Under this circumstance, for Model 1 with the canonical momenta, $p_x$ and $p_y$ as generators, 
we see the following three results which are our main claims of this paper. 
(1) We see a trade-off relation, or an uncertainty relation for joint estimation of the expectation values of two commuting observables.
(2) The trade-off relation, or the uncertainty relation is determined by both of the RLD and the SLD C-R bounds.
Therefore the bound has a non-trivial structure. 
(3) The RLD and SLD C-R bounds have whether no or two intersections 
depending on the fixed expectation value of the angular momentum, $\langle L \rangle_0$. 
Therefore, a transition occurs in the shape of the uncertainty relation depending on the value of $\langle L \rangle_0$. 

For Model 2 with the mechanical (kinetic) momenta, $\pi_x$ and $\pi_y$ as genrators, we see 
another trade-off relation although the observables commute. In contrast, Model 2 
is turned out to be a simple Gaussian shift model 
which was well-studied \cite{holevo,yuen}.  
Therefore, the model is D-invariant and the RLD C-R bound is an achievable bound. 
In the case of the thermal state as the reference state also, Model 2 potentially 
gives a more precise position measurement by estimating the parameters shift generated by the mechanical (kinetic) momentum.
The supplement and the calculations are given in~\ref{sec:suppl} and~\ref{sec:calc}, respectively.  

Throughout the paper, we use the natural units, where we set $c=1$ (the speed of light), 
$\hbar=1$ (the Plank constant), and $k_B=1$ (the Boltzmann constant) unless otherwise stated. 

\section{Preliminaries} \label{sec:prelims}
\subsection{Hamiltonian} \label{sec:phy_model}
The Hamiltonian $H$ for an electron motion in a uniform magnetic field is
\be
H 
= \frac{1}{2m}(\vec{p} + e\vec{A})^2 .  \label{H_pxpy}
\ee
where $-e$ and $m$ are the charge of an electron ($e > 0$), and the mass of the electron, respectively.  
$\vec{A}$ is a vector potential.  In the following discussion, we use the coordinate representation of operators. 
The canonical observables describing this systems are $p_x, \, x, \, p_y,$ and $y$.
We will investigate the uncertainty relation of an electron motion in a uniform magnetic field $\vec{B} = (0, 0, B), \: B > 0$.  
We use the symmetric gauge.  Hence the vector potential is written as $\vec{A} = B \, (-y/2 , x/2 , 0)$. 
We can show that the choice of the gauge gives no change in 
the quantum Fisher information when the magnetic field is uniform. 

We will consider the motion in $x-y$ plane only, because $z$ component solution is a plane wave. 
With a new vector operator, $\vec{\pi} =  \vec{p} + e \vec{A}$, our Hamiltonian becomes~\cite{johnson}   \\[-4mm]
\be
H = \frac{1}{2m}( \pi_x^2 + \pi_y^2).
\label{H_pi}
\ee
Here we remark that these mechanical (kinetic) momenta satisfy the canonical commutation relation up to a constant factor: 
$[ \pi_x,\pi_y\,]= -i eB$ ~\cite{johnson}. They together with the guiding center operators are the fundamental observables in the study of 
electrons in strong magnetic fields, see for example \cite{kubo65}.

It is known that the operators $x,\:y$ and  $p_x, \: p_y$ are equally described by the two sets of 
the creation and annihilation operators, acting on the different Fock spaces, 
$a, a^\dagger$ and $b, b^\dagger$ such that $[a,\, a^\dagger] = [b,\, b^\dagger] = 1$ 
with all other commutation relations vanishing \cite{malkin}. 

The canonical momenta $p_x, \, p_y$ and the position $x, \, y$ in Eq.~\eqref{H_pxpy} are expressed as
\begin{align}
p_x =  \frac{\I}{2\lambda} \left[ ( a^\dagger - a ) + ( b^\dagger - b ) \right],  &\quad 
p_y =  \frac{1}{2\lambda} \left[ ( a^\dagger + a ) - ( b^\dagger + b ) \right], \label{p}\\
x =  \frac{\lambda}{2} \left[ ( a^\dagger + a ) + ( b^\dagger + b ) \right],  &\quad
y = - \frac{\I \lambda}{2} \left[ ( a^\dagger - a ) - ( b^\dagger - b ) \right]. \label{xy} 
\end{align}

The mechanical momenta $\pi_x, \, \pi_y$ in Eq.~\eqref{H_pi} are expressed as 
\begin{align}
\pi_x =  \frac{\I}{\lambda} ( a^\dagger - a ), &\quad \pi_y = \frac{1}{\lambda} ( a^\dagger + a ). \label{pi}\\
\end{align}
where $\lambda = \sqrt{2(eB)^{-1}}$ 
has the dimension of length.  As shown in Eq.~\eqref{LLL2} below, 
$\lambda$ corresponds to the spread of the probability density of the electron in the LLL.

The Hamiltonian $H$ and $z$ component of the angular momentum $L$ are 
expressed in terms of the two harmonic oscillators as
\begin{align}
H &= \omega( a^\dagger a + \frac{1}{2}), \label{H_a} \\
L &= x p_y - y p_x = a^\dagger a - b^\dagger b, \label{Lz}
\end{align}
where $\omega=eB / m$ is the cyclotron frequency. 

\subsection{States} \label{sec:degeneracy}
As the states on which the operators $a, \, a^\dagger$ and $b, \, b^\dagger$ act, the number states $\ket{n}_a$ 
and $\ket{n}_b$ that satisfy 
\be
a^\dagger a \ket{n}_a = n \ket{n}_a, \quad  b^\dagger b \ket{n}_b = n \ket{n}_b,  \label{number}
\ee
are often used.  
The number states $\ket{0}_a$ and $\ket{0}_b$ are the vacuum states of the harmonic oscillators.

Since the Hamiltonian $H$ does not include $b, \, b^\dagger$, its energy eigenstate consists of  
infinite number of the angular momentum eigenstates Eq.~\eqref{Lz}, i.e, the energy eigenstate is degenerated.
We choose the state with the energy $ \omega / 2$ and with zero angular momentum as the reference state.  
This state is written as 
$\ket{0, \, 0} :=  \ket{0}_a\ket{0}_b$ from Eqs.~(\ref{H_a},~\ref{Lz}). 
The wave function of this state is known as the LLL, 
$\psi_{00} ( x, y)$, which is expressed as
\be 
\psi_{00} ( x, y) = \braket{x, \, y \, | \, 0, \, 0} = C \e^{- \frac{x^2 + y^2}{2 \lambda^2}}. \label{LLL}
\ee
where $C$ is the normalization factor. 
Then, the position probability density $ | \psi_{00} ( x, y) |^2$ is
\be
 | \psi_{00} ( x, y) |^2 \propto \e^{- \frac{x^2 + y^2}{ \lambda^2}}. \label{LLL2}
 \ee
This is a Gaussian distribution with its spread $\lambda$ and with its peak at $(x, \, y) = (0, \, 0)$.
\subsection{Estimation of the position} \label{sec:xyMeasQuest}
The unitary transformations of Model 1 and Model 2 make a shift in 
the position probability density of the electron by $\theta = (\thetaone, \, \thetatwo)$.  
From Eqs.~(\ref{shift1},~\ref{shift2}), we have $\thetaone=\langle x \rangle_\theta - \langle x \rangle_0$ and
$\thetatwo=\langle y \rangle_\theta - \langle y \rangle_0$.  
Then, the shifted state from the reference state has a sharp peak at $(x,\,y) = (\thetaone, \, \thetatwo)$. 
Therefore, estimating $\langle x \rangle_\theta$ and $\langle y \rangle_\theta$ is equivalent to infer the shift parameters $\theta = (\thetaone, \, \thetatwo)$. 
(Under the assumption that we know in advance the expectation value of the position operators with respect to the reference state $\rho_0$.)
We estimate the unknown parameters $\thetaone$ and $\thetatwo$ by making arbitrary measurement, 
which is unbiased. 
We then infer the two parameters from the measurement result. 
We shall use the MSE matrix to measure the estimation accuracy of the position of the electron. 
\subsection{Uncertainty relation by quantum C-R inequality} \label{sec:URQuest}
In order to derive the uncertainty relation from the MSE matrix for inferring the position of the electron based on the quantum estimation theory, 
the following two factors are essential to formulate the problem: 
i) Choice of the reference state and 
ii) generators for the shift in the position of the election.
In this paper, we first consider a pure reference state, which is the vacuum state of the two harmonic oscillator. 
Physically, this state is the energy ground state with zero angular momentum. We then consider a mixed reference state affected by 
the thermal noise. 
For the generators of unitary transformations, the most natural choice is the canonical momenta $p_x,\,p_y$. 
We call a parametric family of the states generated by them as Model 1 [Eq.~\eqref{c_ope11}]. The other choice of the generator is 
the mechanical momenta $\pi_x,\,\pi_y$, and we call this family as Model 2 [Eq.~\eqref{c_ope22}]. 

We next derive the uncertainty relation from the quantum C-R inequality.
Consider a general two-parameter model 
of which quantum Fisher information matrix 
is $G_\theta$.  
The quantum C-R inequality then, bounds the MSE matrix $V_\theta = [V_{ij}]$ as ${V_\theta \geq (G_\theta)^{-1}}$. 
In \ref{sec:QCR}, we derive the {following inequalities:} 
\begin{align}
&V_{ 11} - {g_\theta}^{11} \geq 0, \: \: V_{ 22} - {g_\theta}^{22} \geq 0, \label{SLD_CR1} \\
&(V_{ 11} - {g_\theta}^{11})(V_{ 22} - {g_\theta}^{22}) \geq | \, \mathrm{Im} \, {g_\theta}^{12} \, |^2 . \label{CR_ineq}
\end{align}
where $  (G_\theta)^{-1} = [{g_\theta}^{ij }]$. 
We regard these inequalities as a trade-off relation, or an {\it uncertainty relation} for estimating the two parameters 
$\theta=(\thetaone,\,\thetatwo)$. 
In contrast to the Heisenberg-Robertson type uncertainty relation, the commutation relationship between two observables 
do not appear explicitly in the expression above. This is why we can derive a non-trivial uncertainty relation for 
estimating the position of the electron in our models with the commuting observables. 

Note that when the imaginary part of the quantum Fisher information matrix vanishes, i.e., 
$\mathrm{Im} \, {g_\theta}^{12} = 0$, 
the uncertainty relation is given by Eq.~\eqref{SLD_CR1} only.  
In this case, we do not have any trade-off relation between $V_{11}$ and $V_{22}$. 

In the remaining of the paper, we consider the SLD and the RLD quantum Fisher information matrices. 
But our formulation can be extended to any quantum Fisher information matrices. 
\subsection{RLD and SLD Fisher information matrices, generalized RLD information matrix, Z matrix} \label{sec:matrices}
\subsubsection{RLD $L_{\mathrm{R}, \, i}(\theta)$ and RLD Fisher information matrix: $G_\mathrm{R}(\theta)$}
\noindent RLD $L_{\mathrm{R}, \, i}(\theta)$ is given as a solution of the equation below if one exists.
\be
\frac{\partial \rho_\theta}{\partial \thetai} = \rho_\theta L_{\mathrm{R}, \, i}(\theta).  \nonumber 
\ee
The RLD Fisher information matrix $G_\mathrm{R} (\theta) = [g_{\mathrm{R}, \: i j}(\theta)]$ is {defined by}
\be
g_{\mathrm{R}, \: i j}(\theta)= \tr \, [ \rho_\theta \, L_{\mathrm{R}, \, j}(\theta) L^\dagger_{\mathrm{R}, \, i}(\theta) ]. \label{RLD_Fisher}
\ee
\subsubsection{SLD $L_{\mathrm{S}, \, i}(\theta)$ and SLD Fisher information matrix: $G_\mathrm{S} (\theta)$}
\noindent SLD,  $L_{\mathrm{S}, \, i}(\theta)$ is also given as a solution of the equation below if one exists.
\be
\frac{\partial \rho_\theta}{\partial \thetai} = \frac{1}{2} [ \rho_\theta \, L_{\mathrm{S}, \, i}(\theta) 
+  L_{\mathrm{S}, \, i}(\theta) \, \rho_\theta ].
\ee
SLD Fisher information matrix $G_\mathrm{S} (\theta) = [g_{\mathrm{S}, \: i j}(\theta)]$ is defined by
\be
g_{\mathrm{S}, \: i j}(\theta)= \mathrm{Re} \, \tr \, [ \rho_\theta \, L_{\mathrm{S}, \, j}(\theta) L_{\mathrm{S}, \, i}(\theta) ]. \nonumber
\label{SLD_Fisher}
\ee
\subsubsection{Generalized RLD}
In general, the RLD does not exist when a pure state is the reference state~\cite{fujiwara2}.  
We can show that this holds 
for Model 1 and Model 2.  Instead of the RLD Fisher information matrix, 
we are able to obtain the generalized RLD Fisher information matrix by the method introduced by~\cite{fujiwara2}.
Let the generalized RLD Fisher information matrix $\tilde{G}_\mathrm{R}$ be
\be
\tilde{G}_\mathrm{R} = [\tilde{g}_{\mathrm{R}, \, ij}], \nonumber 
\ee
Then,
\be
\tilde{g}_{\mathrm{R}, \, ij}= 4 (\braket{\partial_i \psi | \partial_j \psi} + \braket{\psi | \partial_i \psi} \braket{\psi | \partial_j \psi} ) 
\label{geneRLD}
\ee
\subsubsection{Z matrix}
\noindent ${L_\mathrm{S}}^{ i}(\theta)$ is defined by 
\be
{L_\mathrm{S}}^{ i}(\theta)
= \sum_j {g_\mathrm{S}}^{j i}(\theta) L_{\mathrm{S}, \, j}(\theta). \nonumber
\ee
where $G_\mathrm{S}^{-1} (\theta) = [{g_\mathrm{S}}^{ i j}(\theta)]$.
Then, $Z$ matrix, $Z(\theta) = [ z^{i j}(\theta)]$ is defined by
\be
z^{i j}(\theta)
= \tr \, [\rho_\theta L_\mathrm{S}^{\: j}(\theta) L_\mathrm{S}^{ i \, \dagger}(\theta)] . \nonumber
\ee
It is worth noting the relationship between $Z$ matrix and the expectation value of the commutator of SLD's, 
$\tr (\rho_0 [L_{\mathrm{S}, \, i}(\theta), \, L_{\mathrm{S}, \, j}(\theta)])$ \cite{suzuki}. 
By using the $(i, j)$ component of the $Z$ matrix, $z^{ij}$, we can write the expectation value of the commutator 
$[L_\mathrm{S}^{ i}(\theta), \, L_\mathrm{S}^{ j}(\theta)]$ as
\be
\tr(\rho_0 [ L_{\mathrm{S},i} , \, L_{\mathrm{S},j}]) =\sum_{k,\ell}g_{\mathrm{S},ki} \left(z^{k\ell} - (z^{k\ell})^\ast\right) g_{\mathrm{S},\ell j} 
= 2 \I  \, \sum_{k,\ell}g_{\mathrm{S},ki} \mathrm{Im} (z^{k\ell}) g_{\mathrm{S},\ell j}. 
\ee 
In particular, the expectation value of the commutator is proportional to the imaginary part of $Z$ matrix when the SLD Fisher information matrix is diagonal. 
%
\section{Pure state model} \label{sec:Pure_st}
In this section, we first consider an ideal situation, where the reference states are given by a pure state. 
The derived uncertainty relation for Model 1 is understood intuitively, since the model is two-independent unitary model. 
Model 2, which is generated by the two non-commuting generators, gives a non-trivial uncertainty relation. 
\subsubsection{Reference state}
Since the energy eigenstate of Hamiltonian \eqref{H_pi} is infinitely degenerated, we choose the tensor product of the vacuum states 
as the reference state $\rho_0$ which is denoted by 
\\[-5mm]
\begin{align}
\rho_0 &= | 0 \rangle_a \, {_a} \langle 0 \, |  \otimes | 0 \rangle_b \, {_b} \langle 0 \, | = \ket{0, \, 0} \bra{0, \, 0}. 
\label{ref_p}
\end{align}

\subsubsection{Unitary transformations}
We introduce two kinds of unitary transformations,  $\e^{- \I \thetaone p_x}  \e^{- \I \thetatwo p_y}$ and 
$\e^{- \I \thetaone \pi_x}  \e^{- \I \thetatwo \pi_y}$.  We consider that we have them act on the LLL, $\psi_{00} (x, y)$.
Since we have
\be
\e^{- \I \thetaone p_x}  \e^{- \I \thetatwo p_y}  \psi_{00} (x, y)  =  \psi_{00}(x - \thetaone, y - \thetatwo), \label{trans_p}
\ee
this unitary transformation $\e^{- \I \thetaone p_x}  \e^{- \I \thetatwo p_y}$ makes a shift in $x-y$ coordinate of
$\psi_{00} (x, y)$ from $(x, \, y)$ to $(x - \thetaone, y - \thetatwo)$.  
We also have 
\be
\e^{- \I \thetaone \pi_x}  \e^{- \I \thetatwo \pi_y} \psi_{00} (x, y) 
= \e^{ \I \frac{\thetaone \thetatwo }{\lambda^2}} \e^{- \I \frac{x \thetatwo }{\lambda^2}} \e^{ \I \frac{ y \thetaone}{\lambda^2}} 
 \psi_{00}(x - \thetaone, y - \thetatwo) , \label{trans_Pi}
\ee
where we use the standard Baker-Campbell-Hausdorff formula \cite{klauder}. 
Because the difference between Eqs.~(\ref{trans_p},~\ref{trans_Pi}) is only in the phase shift of 
the wave function, 
the unitary transformations $\e^{- \I \thetaone \pi_x}  \e^{- \I \thetatwo \pi_y}$ and 
$\e^{- \I \thetaone p_x}  \e^{- \I \thetatwo p_y}$ give the same effect to the probability density, $ | \psi_{00}(x, y) |^2 $, i.e., both of them make the shift as follows.
\begin{align}
 | \e^{- \I \thetaone p_x}  \e^{- \I \thetatwo p_y} \psi_{00}(x, y) |^2 
 &=  | \psi_{00}(x - \thetaone, y - \thetatwo) |^2, \nonumber   \\
  | \e^{- \I \thetaone \pi_x}  \e^{- \I \thetatwo \pi_y} \psi_{00}(x, y) |^2 
  &=  | \psi_{00}(x - \thetaone, y - \thetatwo) |^2 . \label{sh2} 
\end{align}
and thus, we have in both cases,
\be
 | \psi_{00}(x - \thetaone, y - \thetatwo) |^2 \propto \exp\left[- \frac{(x - \thetaone)^2 + (y - \thetatwo)^2 }{\lambda^2}\right].
 \ee
 That is, the position probability density is now centered at $(\thetaone,\,\thetatwo)$ with its spread $\lambda$. 
\subsection{Uncertainty relation}

It is known that the RLD does not exist in general when the reference state is a pure state. 
In Ref.~\cite{fujiwara}, the authors showed that the SLD can be defined by taking equivalent classes of 
the inner product, thereby the SLD Fisher information matrix exists uniquely. 
In later work \cite{fujiwara2}, they also showed that the generalized RLD Fisher information matrix exists 
for a special class of pure-state models, called a coherent model. In our case, 
Model 1 turns out to be a quasi-classical model meaning that the SLD Fisher information matrix 
plays the same role as the classical Fisher information matrix. 
Whereas Model 2 is shown to be a 
coherent model, and hence we can derive the quantum C-R inequality 
based on the generalized RLD Fisher information matrix. 
\subsubsection{Model 1: Unitary model generated by $p_x$ and $p_y$} \label{sec:Pure_st1}
In Model 1, the generators, $p_x$ and $p_y$ commute.  
We can show that the SLD's commute on the support of the states and that the SLD 
C-R bound is achievable~\cite{matsumoto}. 

The SLD and the generalized RLD Fisher information matrices are calculated by the way given in \cite{fujiwara}.
Since the Fisher information matrices of the unitary models do not depend on $\theta$, we omit $\theta$ for simplicity. 
The SLD Fisher information matrix with respect to the reference state $\rho_0$ is denoted by $\Gspp$. 
Then, its inverse is calculated in \ref{sec:gen_RLD1} as 
\be
(\Gspp )^{-1} = \frac{\lambda^2}{2} \begin{pmatrix}
1  & 0 \\
0 & 1  \\
\end{pmatrix},   \nonumber
\ee

From Eq.~\eqref{SLD_CR1}, we obtain 
\be
V_{11} \geq \frac{\lambda^2}{2} , \: \: 
V_{22} \geq \frac{\lambda^2}{2} . \nonumber
\ee

We next calculate the the generalized RLD Fisher information matrix $\GGrpp$ 
to find out that the off-diagonal components of $\tilde{G}_\mathrm{R}^p$ are zero and that $G^p_\mathrm{S}=\tilde{G}_\mathrm{R}^p$ for Model 1. 
Then, $(G^p_\mathrm{S})^{-1}=(\tilde{G}_\mathrm{R}^p)^{-1}=Z$ holds even though this model is not coherent. 
As is explained in the following section, we thus use the generalized RLD for Model 2 only.
The calculation is shown in \ref{sec:gen_RLD1}. 
Figure~\ref{fig4} shows the SLD C-R bound (dotted lines).
${\lambda^2} / {2}$ is a half of the square of the spread of the LLL wave function
Eq.~\eqref{LLL2}.  This result shows that the measurement accuracy is limited by the spread of 
the probability density of the electron in the LLL.  This results from the quasi-classical nature of 
Model 1. 
\subsubsection{Model 2: Unitary model generated by $\pi_x$ and $\pi_y$} \label{sec:Pure_st2}
Let $\Gspip$ denote the SLD Fisher information matrix of Model 2 with respect to the reference state $\rho_0$. 
Then, the inverse of SLD Fisher information matrix $(\Gspip )^{-1}$ is calculated in~\ref{sec:gen_RLD2} as\\[-3mm]
\be
(\Gspip )^{-1}= \frac{\lambda^2}{4} \begin{pmatrix}
1  & 0 \\
0 & 1  \\
\end{pmatrix}. \nonumber
\ee 
Notably, the relation $(\Gspp )^{-1} = 2 (\Gspip )^{-1}$ holds.  
This difference, a factor of two results from the difference in the coefficients in Eqs.~(\ref{p},~\ref{pi}).

Let $\GGrpip$ denote the  generalized RLD Fisher information matrix.  
The generalized RLD C-R bound \cite{fujiwara2} is given by ${V_\theta \geq (\GGrpip)^{-1}}$, where
\be
(\GGrpip )^{-1}= \frac{\lambda^2}{4} \begin{pmatrix}
1  & \I \\
-\I & 1  \\
\end{pmatrix}. \label{GRLD}
\ee
The derivation of~\eqref{GRLD} is given in~\ref{sec:gen_RLD}. 
By using Eq.~\eqref{CR_ineq}, we obtain the following inequality, 
\be
(V_{11} - \frac{\lambda^2}{4})(V_{22} - \frac{\lambda^2}{4}) \geq \frac{\lambda^4}{16}. \label{M2_p_bnd}
\ee

\begin{figure}[t]
\begin{center}
\includegraphics[width=8cm]{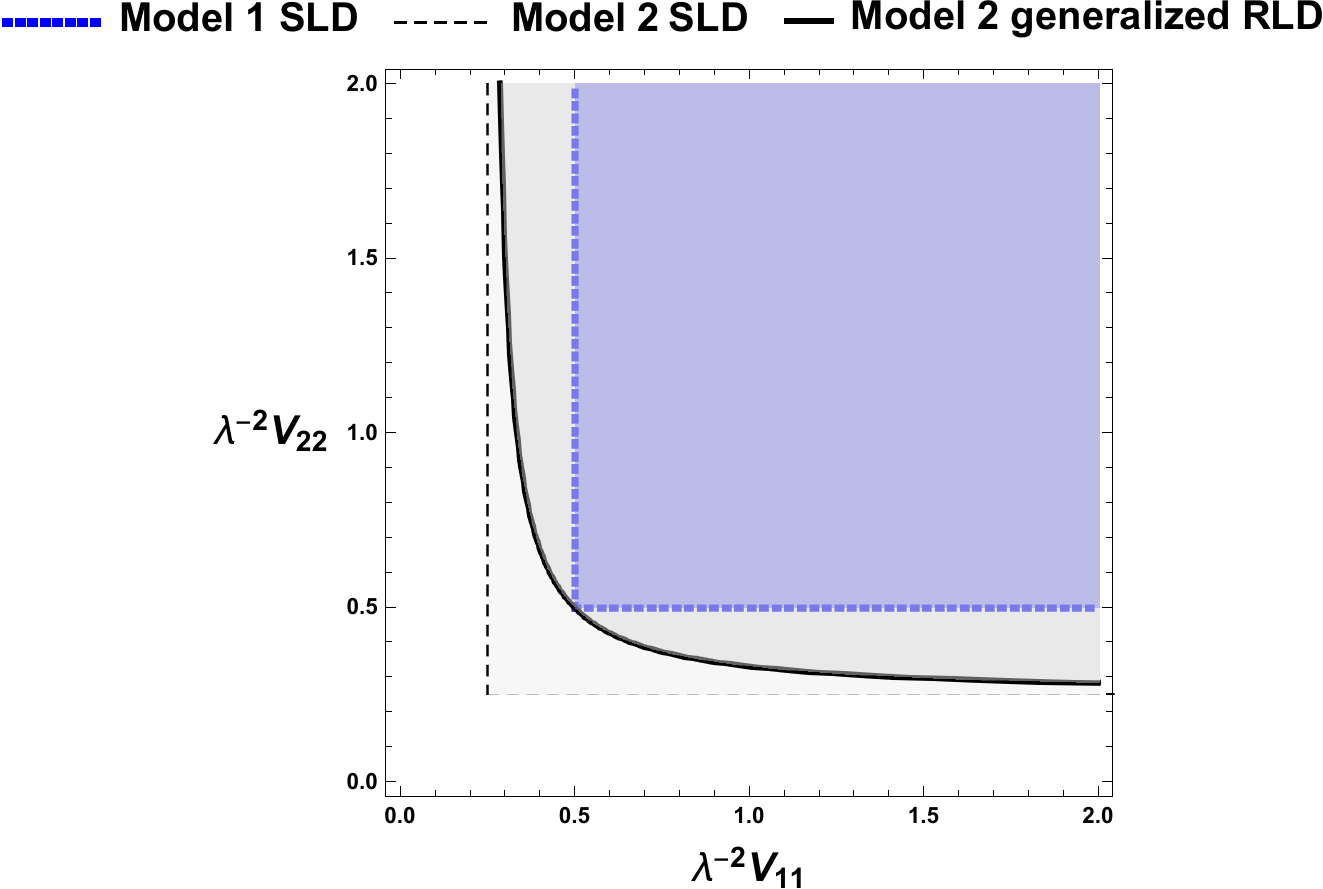}
\caption{The uncertainty relation of Model 1 and Model 2 given by the inequalities Eqs.~(\ref{SLD_CR1},~\ref{CR_ineq}). 
The allowed region of Model 2 for the MSE matrix components $(V_{11}, \,  V_{22})$ is above the solid line, 
the dark gray region and the blue region 
given by the inequality~\eqref{M2_p_bnd} which is derived from the generalized RLD Cram\'er-Rao inequality.  The allowed region 
of Model 2 by the SLD C-R inequality consists of the light gray, dark gray and blue regions.
The allowed region 
of Model 2 by the SLD Cram\'er-Rao inequality is the blue region.  }
\label{fig4}
\end{center}
\end{figure}

Figure~\ref{fig4} shows the SLD C-R bound (dashed line) and the generalized RLD C-R bound (solid line). 
Since the generators, $\pi_x$ and $\pi_y$ consist of $a, \, a^\dagger$ only [Eq.~\eqref{pi}], this is a 
Gaussian shift model and the generalized RLD bound is achievable \cite{fujiwara2}.
Unlike the result of Model 1, 
the uncertainty relation for Model 2 exhibits a trade-off relation between $V_{11}$ and $V_{22}$. 
This comes from the nature of Model 2 which is purely quantum mechanical. 

\subsection{Discussion}
There are two significant differences between the C-R bounds given by Model 1 and Model 2 
even though the unitary transformations of Model 1 and Model 2 make the same shift in the position 
of the probability density as shown in Eq.~\eqref{sh2}.

First, the SLD and the generalized RLD C-R bounds of Model 2 is lower than the SLD C-R bound of Model 1. 
In particular, the SLD C-R bound of Model 2 is a half of that of Model 1.  
At first sight, this difference in estimation accuracy might puzzle us, since two models displace the same amount in the position. 
However, there is no inconsistency in our models, and the simple answer is given as follows. 
Note that the generators for Model 2 shift twice of Model 1 as in Eqs. (\ref{p}, \ref{pi}) in the parameter space.   
This results in the larger quantum Fisher information of Model 2.  
The measurement accuracy appears to be better in Model 2 only because Model 2 shifts more. 

Second, Eq.~\eqref{M2_p_bnd} gives the achievable bound of {Model 2} \cite{fujiwara2}.  
The relation between $V_{1 1}$ and $V_{2 2}$ in the right hand side of Eq.~\eqref{M2_p_bnd} 
is not just a product of $V_{1 1}$ and $V_{2 2}$ 
unlike the Heisenberg-Robertson type uncertainty relation. 
The difference in the bound between Model 1 and Model 2 is caused by  
the phase shift of Model 2 in Eq.~\eqref{trans_Pi}.  
Although this phase shift in Eq.~\eqref{trans_Pi} makes no change in the position probability density, 
it does make a change in the quantum Fisher information matrices.  
\section{Mixed state model: Effect of thermal noise} \label{sec:Mixed_st}
Next, we use a mixed state as the reference state 
to see how the noise affects the measurement accuracy of the electron position.  
For this purpose, as the mixed state, we choose the thermal state.  However, in the current system we are considering, 
there is no unique thermal state, because the energy eigenstate is degenerated. 
Then, the thermal state of this system is not uniquely specified by the temperature only.  
To resolve this degeneracy problem, we impose 
a condition that the expectation value of the angular momentum $\langle L \rangle_0$ is fixed.  This is done 
by introducing a chemical potential.
\subsection{Reference state}
Given $\langle L \rangle_0$ is fixed at a constant, 
the reference state $\rho_{\beta, \, \mu}$ is denoted by 
\be
\rho_{\beta, \, \mu}
= Z_{\beta, \, \mu}^{-1} \e^{-\beta H + \mu L}, 
\ee
where $\beta=T^{-1}$ is the inverse temperature and $Z_{\beta, \, \mu} = \tr \,[ \exp(-\beta H + \mu L) ]$ is the partition function. 
The parameter $\mu$ is the chemical potential, which will be determined later. 
The role of the chemical potential $\mu$ is to keep $\langle L \rangle_0$ constant to 
avoid complications by the degeneracy of angular momentum.
The use of the chemical potential here is the same idea as seen in the grand canonical ensemble of statistical physics  
where the chemical potential is used to keep the expectation value of the number of particles constant. 

From Eqs.~(\ref{H_a},~\ref{Lz}), 
\be
\rho_{\beta, \, \mu} = Z_{\beta, \, \mu}^{-1} 
\e^{- \frac{1}{2} \beta \omega} \e^{-(\beta \omega - \mu) a^\dagger a  - \mu b^\dagger b}.
\ee
By using the Gaussian states which are defined by 
\be
a \ket{z}_a = z \ket{z}_a, \: \: b \ket{z}_b = z \ket{z}_b,
\ee
the reference state $\rho_{\beta, \, \mu}$ is expressed as
\be
\rho_{\beta, \, \mu} = \rho_{0, \, a} \otimes  \rho_{0, \, b}, \label{thermal_ab}
\ee
where $\rho_{0, \, a}$ and $\rho_{0, \, b}$ are the thermal states with different temperatures. 
Explicitly, they are 
\begin{align}
\rho_{0, \, a} &= \frac{1}{2 \pi \ka^2} \int \e^{- \frac{| z |^2}{2 \ka^2}} \, | z \rangle_a \, {_a} \langle z \, | \, d^2 z, 
\nonumber   \\
\rho_{0, \, b} &=\frac{1}{2 \pi \kb^2} \int \e^{- \frac{| z |^2}{2 \kb^2}} \, | z \rangle_b \, {_b}
\langle z \, | \, d^2 z,  \nonumber
\end{align} 
with
\be
2 \ka^2 = (\e^{  \beta \omega - \mu}-1)^{-1}, \quad
2 \kb^2 = (\e^{\mu} - 1)^{-1}.  \label{kappas}
\ee
The derivation of Eqs.~(\ref{thermal_ab},~\ref{kappas}) is given in~\ref{sec:thermal_state}. 
It is straightforward to calculate the expectation value $\langle L \rangle_0 $ as
\be
\langle L \rangle_0 = \tr \, [ L \, \rho_{\beta, \, \mu}] 
= 2 \ka^2 - 2 \kb^2. \label{Lz_kappa}
\ee

From Eqs.~(\ref{kappas},~\ref{Lz_kappa}), we obtain 
\be
(\langle L \rangle_0+1) \e^{2\mu}- \langle L \rangle_0 (\e^{\, \beta \omega} +1) \e^{\mu}+(\langle L \rangle_0-1)\e^{\, \beta \omega}= 0. \label{L_eq}
\ee
When $\beta \omega$ and $\langle L \rangle_0 $ are given, $\mu$ 
is the variable of Eq.~\eqref{L_eq}.   
If $\langle L \rangle_0=-1$ holds, there exists a unique solution. Whereas there are two solutions for $\langle L \rangle_0\neq-1$.  
However, one of them is shown to be unphysical giving a negative temperature state in the later case. 
Then, the chemical potential $\mu$ as a function of $\langle L \rangle_0$ and $\beta \omega$ is found to be 
\begin{equation} \label{mu}
\e^{\mu}=\begin{cases}
\displaystyle\frac{2 \e^{\, \beta \omega} }{\e^{\, \beta \omega}+1} \hfill (\langle L \rangle_0 = -1) \\[1.5ex]
\displaystyle \frac{1}{2(\langle L \rangle_0+1)}\Big[\langle L \rangle_0 ( \e^{\, \beta \omega}+1) 
+\sqrt{ \langle L \rangle_0^2 (\e^{\, \beta \omega}-1)^2+4\e^{\, \beta \omega}  } \Big] \\
   \hfill  (\langle L \rangle_0 \neq -1) 
\end{cases} .  
\end{equation}
Although the solution of Eq.~\eqref{L_eq} has a singular point at $\langle L \rangle_0 = -1$ at a first glance, 
we can show that the solution the solution for $\langle L \rangle_0 \neq -1$ is 
continuously connected to the solution for $\langle L \rangle_0 = -1$.  We can also show that the first derivative is 
continuous at $\langle L \rangle_0 = -1$.

\begin{figure}[t]
\begin{center}
\includegraphics[width=7cm]{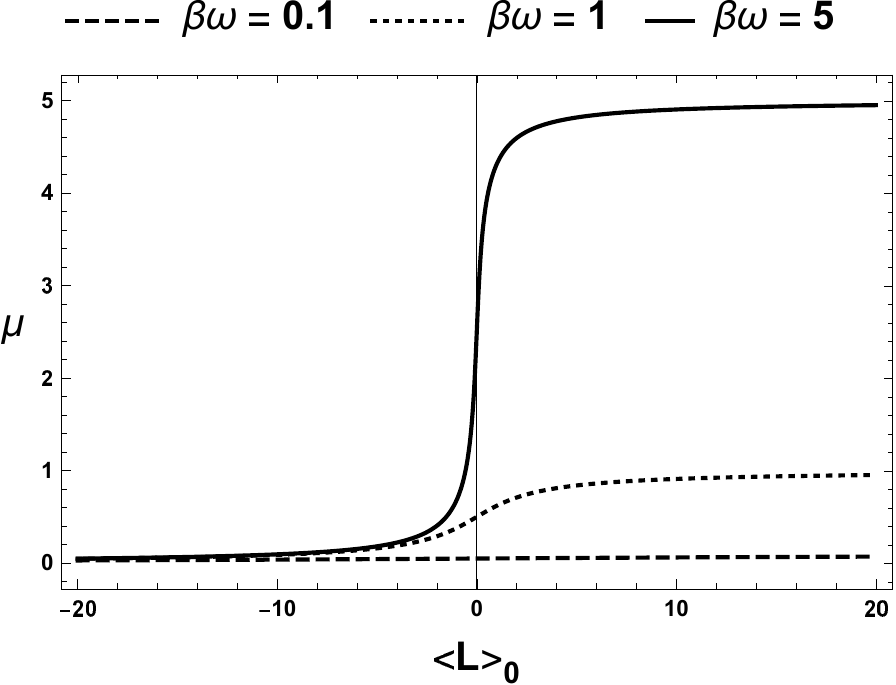}
\caption{The chemical potential $\mu$ as a function of the expectation value of the angular momentum 
$\langle L \rangle_0$ at three different temperature parameters $\beta \omega = 0.1, \, 1$, and $5$. 
At lower $\beta \omega$ i.e., higher temperature, $\mu$ becomes closer to zero, 
no preference for the angular momentum.}
\label{fig1}
\end{center}
\end{figure}

Figure~\ref{fig1} shows $\mu$ as a function of $\langle L \rangle_0$ at $\beta \omega = 0.1, \, 1,$ and $5$ from top to bottom.  
The chemical potential $\mu$ as a function of $\langle L \rangle_0$ diverges for $\langle L \rangle_0\ge0$ 
as $\beta \omega$ goes to infinity, 
i.e., the zero temperature limit. 
At a special case, $\langle L \rangle_0 = 0$, we see $\mu = \beta \omega / 2$ from Eq.~\eqref{mu}. 
Explicitly, the zero temperature limit is 
\begin{equation}
\lim_{\beta\to\infty} \mu =\begin{cases}
\infty&\quad (\langle L \rangle_0 \ge 0)\\[1.5ex]
\log\left[\frac{\langle L \rangle_0-1}{\langle L \rangle_0} \right]&\quad  (\langle L \rangle_0 < 0)
\end{cases} .
\end{equation}

For Model 2, the two-parameter family of the states $\rho_\theta^\pi$ is expressed as
\be
\rho^\pi_\theta= \e^{- \I \thetaone \pi_x} \e^{- \I \thetatwo \pi_y} \rho_{0, \, a} \otimes \rho_{0, \, b} 
\, \e^{ \I \thetatwo \pi_y} \e^{ \I \thetaone \pi_x},
\ee
from Eqs.~(\ref{c_ope22},~\ref{thermal_ab}). 
Since $\pi_x$ and $\pi_y$ consist of $a$ and $a^\dagger$ only, $\rho^\pi$ is described as
\be
\rho^\pi_\theta= \e^{- \I \thetaone \pi_x} \e^{- \I \thetatwo \pi_y} \rho_{0, \, a}  
\, \e^{ \I \thetatwo \pi_y} \e^{ \I \thetaone \pi_x} \otimes \rho_{0, \, b}.
\ee
Therefore, $\rho_{0, \, b}$ gives no effect to the quantum Fisher information matrices.  
The reference state $\rho_{\beta, \, \mu}$ for Model 2, we only need to use $\rho_{0, \, a}$. 
By construction, the family of the states:
\[
\rho_{\theta,\,a}= \e^{ \xi a^\dagger - \xi^\ast a }  \rho_{0, \, a}  
\,  \e^{ \xi^\ast a - \xi a^\dagger } , 
\]
with $\xi =  ( 2 \lambda)^{-1} (\thetaone - \I \thetatwo )$, is a Gaussian shift model \cite{holevo, yuen}. 
It is then known that the RLD C-R bound provides the achievable bound~\cite{holevo, yuen}.
\subsection{Uncertainty relation}
For the mixed-state model, we can calculate the SLD and the RLD C-R bounds. They then provide 
the uncertainty relation for the MSE matrix. The calculations of SLDs and RLDs and their quantum Fisher information matrices are given in~\ref{sec:SLD_RLD}.

\subsubsection{Model 1: Unitary model generated by $p_x$ and $p_y$} 
Let $\Grpt$ and $\Gspt$ be the RLD and the SLD Fisher information matrices with respect to $\rho_{\beta, \, \mu}$, 
respectively. We introduce ${g_\mathrm{R}}^{ij}$ and ${g_\mathrm{S}}^{i j}$ such that 
\begin{align}
(\Grpt)^{-1} &= [ {g_\mathrm{R}}^{ij}],  \\
(\Gspt)^{-1} &= [{g_\mathrm{S}}^{i j}]. 
\end{align}
The inverse of $\Grpt$ is calculated as
\be
(\Grpt)^{-1}
=  \frac{\lambda^2}{ 1 + 2 \ka^2 + 2 \kb^2} 
\begin{pmatrix}
  2 \ka^2 + 2 \kb^2 + 8 \ka^2 \kb^2 & \I   \, (2 \kb^2 - 2 \ka^2) \\
- \I   \, (2 \kb^2 - 2 \ka^2) &   2 \ka^2 + 2 \kb^2 + 8 \ka^2 \kb^2  \\
\end{pmatrix}. \nonumber
\ee

From Eq.~\eqref{CR_ineq}, 
we have the following inequality, 
\be
(V_{11} - {g_\mathrm{R}}^{11} ) (V_{ 22} - {g_\mathrm{R}}^{11} ) 
\geq \lambda^4 \left( \frac{2 \ka^2 - 2 \kb^2}{ 1 + 2 \ka^2 + 2 \kb^2} \right)^2. \label{RLD-CR_M1}
\ee
Next, the calculation of the inverse of $\Gspt$ reveals that $(\Gspt)^{-1}$ is a diagonal matrix and that 
${g_\mathrm{S}}^{11}$ is equal to ${g_\mathrm{S}}^{22}$.  $(\Gspt)^{-1}$ is written as
\be
(\Gspt)^{-1} 
= \begin{pmatrix}
{g_\mathrm{S}}^{11}  & 0 \\
  0 & {g_\mathrm{S}}^{22}   \\
\end{pmatrix},
\label{GS_inv}
\ee 
where
\be
{g_\mathrm{S}}^{11}={g_\mathrm{S}}^{22}
= \lambda^2 \frac{ \frac{1}{2} + 2 \ka^2 + 2 \kb^2 + 8 \ka^2 \kb^2}{ 1 + 2 \ka^2 + 2 \kb^2}. \nonumber
\ee
From Eq.~\eqref{SLD_CR1}, we have
\be
V_{11} \geq {g_\mathrm{S}}^{11},  \quad
V_{22} \geq {g_\mathrm{S}}^{11}.  \label{M2_V22}
\ee
There are two cases regarding the ordering between the inverse of RLD and SLD Fisher matrices 
in terms of the matrix inequality. 

Case i). When $ | \langle L \rangle_0 | \leq 1 / 2$, the SLD C-R bound defines a tighter lower bound. 
This is because the matrix inequality 
\be
(\Gspt)^{-1} - (\Grpt)^{-1} 
=  \Delta g \begin{pmatrix}
1 &  - 2 \I   \, \langle L \rangle_0 \\
2 \I   \, \langle L \rangle_0 &  1  \\
\end{pmatrix} \geq 0, \nonumber
\ee
holds if and only if $| \langle L \rangle_0 | \leq 1 / 2$ is satisfied. 
Here, $\Delta g$ is defined by
\be
\Delta g := {g_\mathrm{S}}^{11}  - {g_\mathrm{R}}^{11} 
= \frac{\lambda^2}{2} \frac{1}{1 + 2\ka^2 + 2 \kb^2} > 0. \label{delta_g} \\
\ee

\begin{figure}[t]
\begin{center}
\includegraphics[width=8cm]{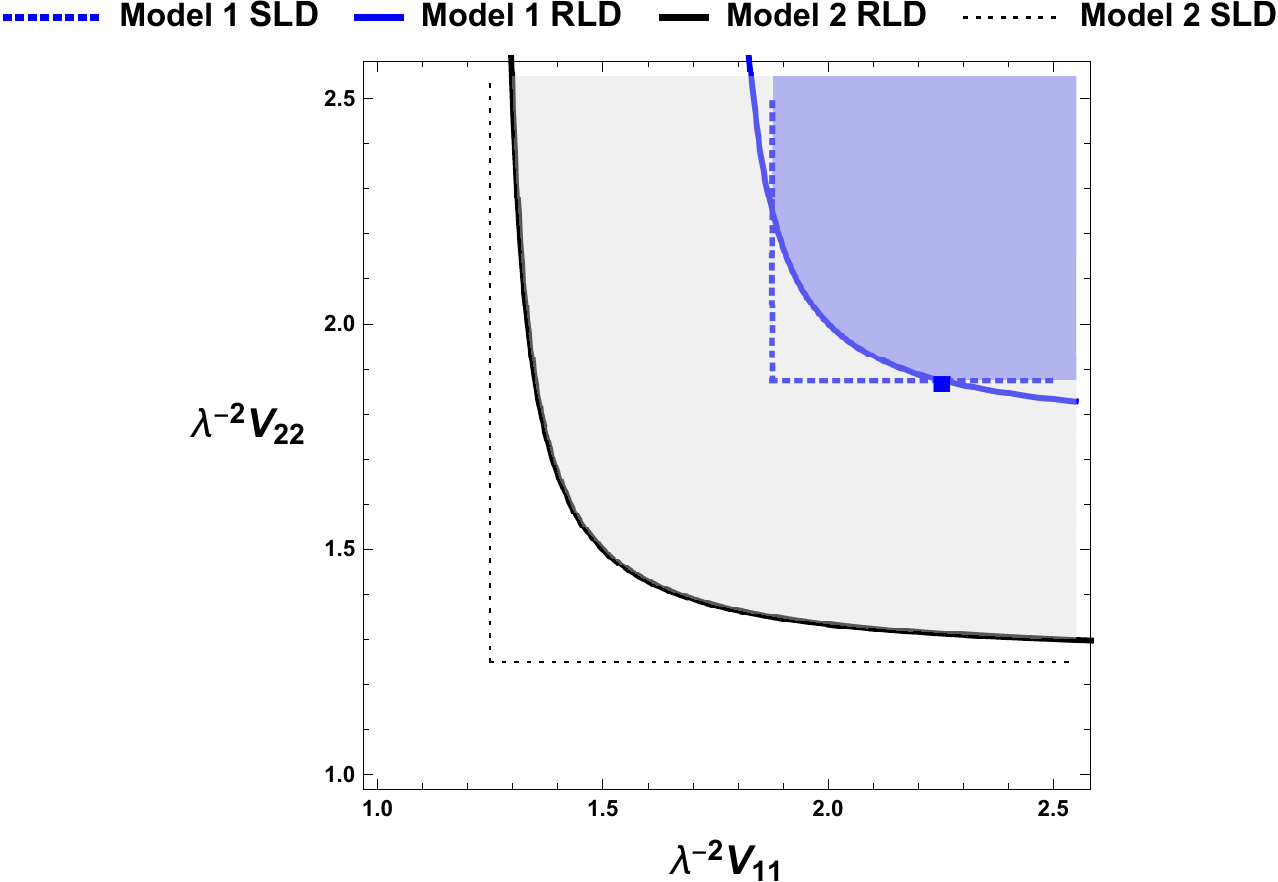}
\caption{Uncertainty relation of Model 1 and Model 2 given by the quantum Cram\'er-Rao inequalites. 
The temperature parameters used are  
$\, \ka^2 = 1, \,  \kb^2 = 1 / 2, \, \langle L \rangle_0 = 1$, i.e., $\langle L \rangle_0 > 1/2$. 
The allowed region of {Model 1} for the MSE matrix components $(V_{1 1}, \,  V_{2 2})$ is the blue region.
The allowed region of Model 1 is given by the SLD Cram\'er-Rao bound (blue dotted lines) and the RLD C-R bound 
(blue solid line).  
The allowed region of Model 2 for the MSE matrix components $(V_{1 1}, \,  V_{2 2})$ is covered by the gray and blue region.
The RLD Cram\'er-Rao bound (black solid line) is achievable. $\Delta V^\mathrm{R-S}$ [Eq.~\eqref{V_RS}] is the distance between the blue square in the figure and the intersection of SLD Cram\'er-Rao bounds (blue dotted lines).}
\label{fig2}
\end{center}
\end{figure}

\begin{figure}[t]
\begin{center}
\includegraphics[width=8cm]{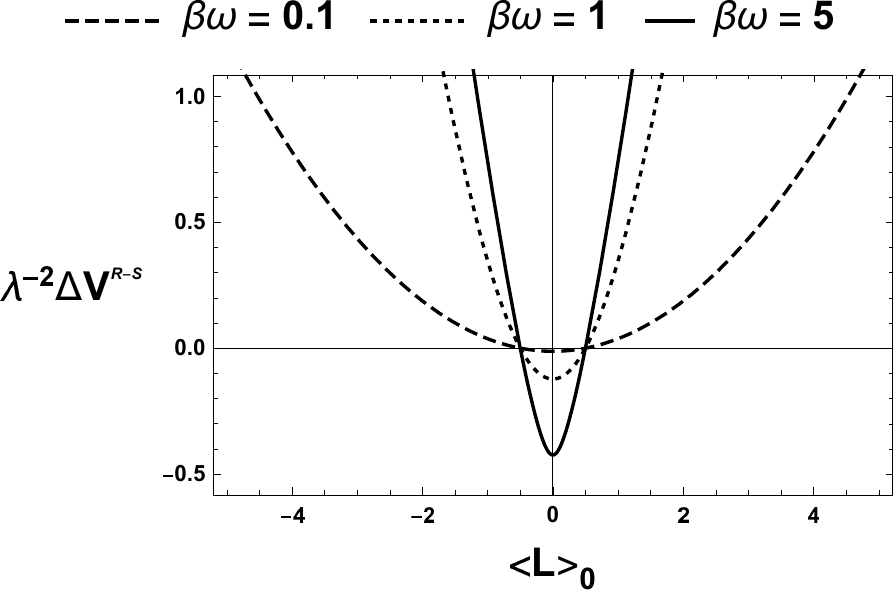}
\caption{$\Delta V^{\mathrm{R-S}}$ [Eq.~\eqref{V_RS}] as a function of $\langle L \rangle_0$. 
By the definition of $\Delta V^{\mathrm{R-S}}$, when $\Delta V^{\mathrm{R-S}}$ is negative, the bound is determined by 
the SLD Cram\'er-Rao bound only.  }
\label{fig3}
\end{center}
\end{figure}

Case ii). In the other case, $| \langle L \rangle_0 | > 1 / 2$, however, 
there is no matrix ordering between the RLD and the SLD Fisher information matrices.  
This means that both inequalities \eqref{RLD-CR_M1} and \eqref{M2_V22} contribute to the uncertainty relation. 
Figure~\ref{fig2} shows an example of the bound given by the current analysis with $| \langle L \rangle_0| > 1 / 2$. 
The parameters used are $ \ka^2 = 1, \,  \kb^2 = 1 / 2$, and thus $| \langle L \rangle_0| = 1 > 1 / 2$ holds.  
The blue region defined by two quantum C-R bounds, the SLD and the RLD C-R bounds are the allowed region of 
 $(V_{11}$,~$V_{22})$.
The RLD and the SLD C-R bounds have two intersection points in this case. 
Let the position of one of the intersection points be $(V^\mathrm{R-S}_{11}, \, {g_\mathrm{S}}^{11})$ 
which is marked as the dot in Fig.~\ref{fig2}.    
The RLD C-R bound defines the bound in the region where ${g_\mathrm{S}}^{11} < V_{11} <  
V^\mathrm{R-S}_{11}$.   
The SLD C-R bound defines in the region where $ V_{11} > V^\mathrm{R-S}_{11}$ and 
$V_{22}  >  V^\mathrm{R-S}_{11}$.  
We define $\Delta V^\mathrm{R-S}$ by $\Delta V^\mathrm{R-S} = V^\mathrm{R-S}_{11} - {g_\mathrm{S}}^{11}$.  
Then, $\Delta V^\mathrm{R-S}$ is
\be
\Delta V^\mathrm{R-S} = \Delta g (4 \langle L \rangle_0^2 - 1). 
\label{V_RS}
\ee
Figure~\ref{fig3} shows $\Delta V^\mathrm{R-S}$ as a function of $\langle L \rangle_0$ at three different $\beta \omega$'s 
which are the same as Fig.~\ref{fig1}.

When $| \langle L \rangle_0 | \le 1 / 2$, $\Delta V^\mathrm{R-S}$ is negative as shown in Eq.~\eqref{V_RS}, the RLD C-R bound stays always below the {SLD C-R} bound.  
This is consistent with $(G_\mathrm{S})^{-1} \geq (G_\mathrm{R})^{-1}$ when $| \langle L \rangle_0 |  \leq 1 / 2$. At larger $\beta \omega$ (lower temperature), the possible ranges of $V^\mathrm{R-S}_{11}$ and $V^\mathrm{R-S}_{22}$ given by the RLD 
C-R bound becomes larger at the same $\langle L \rangle_0$.

Finally, we briefly discuss achievability of the uncertainty relation above. 
It is known that the RLD C-R bound is (asymptotically) achievable, 
if and only when the model is D-invariant \cite{suzuki}. 
This condition is checked by comparing two matrices, the inverse of the 
RLD Fisher information matrix and the $Z$ matrix. 
As given in~\ref{sec:calc}, $(\Grpt)^{-1}$ and the $Z$ matrix $Z^{p \, \mathrm{thermal}}$ are different. 
Hence, the RLD C-R bound is not tight. 
We next examine if the SLD C-R bound is achievable or not. 
In Refs.~\cite{RJDD16,suzuki18}, the necessary and sufficient conditions are derived 
for asymptotically achievability of the SLD C-R bound. 
The simplest condition is that the imaginary part of the $Z$ matrix is zero. 
In our model, this is equivalent to $\langle L \rangle_0 = 0$ which is also equivalent to $\ka = \kb$ [Eq.~\eqref{Lz_kappa}].
When $\langle L \rangle_0 \neq0$, neither the RLD C-R bound nor SLD C-R bound is even asymptotically achievable.  
Therefore, the uncertainty relation is not tight, except for the special choice of the parameter, 
$\langle L \rangle_0 = 0$.

\subsubsection{Model 2 : Unitary model generated by $\pi_x$\,and\,$\pi_y$}

The SLD and the RLD Fisher information matrices of {Model 2} are denoted 
by $\Gspit$ and $\Grpit$, respectively.   
Their inverse matrices $(\Gspit )^{-1}$ and  $(\Grpit )^{-1}$ are \\[-3mm]
\begin{align}
(\Gspit)^{-1}
&= \frac{\lambda^2}{4} \begin{pmatrix}
1 + 4 \ka^2  & 0 \\
0 & 1 + 4 \ka^2  \\
\end{pmatrix} ,\\
(\Grpit )^{-1}
&= \frac{\lambda^2}{4} \begin{pmatrix}
1 + 4 \ka^2  & \I \\
- \I & 1 + 4 \ka^2  \\
\end{pmatrix} .
\end{align}
Since this model is a Gaussian shift model \cite{holevo, yuen}, 
the RLD C-R bound is achievable. 
By using Eq.~\eqref{CR_ineq}, the RLD C-R inequality gives the following inequality
\be
\left[ V_{ 11} - \frac{\lambda^2}{4} (1 + 4 \ka^2 )\right] \left[ V_{ 22} - \frac{\lambda^2}{4}(1 + 4 \ka^2) \right] \geq \frac{\lambda^4}{16}.
\label{RLD_M1}
\ee

From Eq.~\eqref{SLD_CR1}, we obtain the SLD C-R bound as follows.
\be
V_{ 11} \geq \frac{\lambda^2}{4} (1 + 4 \ka^2 ) , \: \: 
V_{ 22} \geq \frac{\lambda^2}{4} (1 + 4 \ka^2 ). \nonumber 
\ee
Figure~\ref{fig2} shows the RLD C-R bound and the SLD C-R bound above
for the temperature parameter $ \ka^2 = 1$ as well.  The gray region is 
the uncertainty relation given by the RLD C-R bound. 
The SLD and RLD C-R bounds move away from the origin$= (0, 0)$ as $\ka^2$ increases.  
This makes sense, because the increase in $ \ka^2$ means the decrease in $\beta $ because  
$2 \ka^2 = (\e^{  \beta \omega - \mu}-1)^{-1}$.  
The C-R bounds of Model 2 stays lower than that of Model 1.
%
\subsection{Discussion}\label{sec:4} 
\subsubsection{Mixed state model}
It turns out that in the case of the thermal state as the reference state, Model 2 is 
a simple Gaussian shift model which is known to be the RLD C-R inequality giving 
an achievable bound \cite{holevo,yuen}. 
However, the bound for Model 1 has a complicated structure as shown in Fig.~\ref{fig2}, although the bound for 
the pure state is simple.  As given in~\ref{sec:alt_rep}, the two-parameter unitary transformation for Model 1, 
$\e^{- \I \thetaone p_x}  \e^{- \I \thetatwo p_y}$ can be written as
\be 
\e^{- \I \thetaone p_x}  \e^{- \I \thetatwo p_y} =  \e^{ \xi_1 a^\dagger - \xi_1^\ast a }  \e^{ \xi_2^\ast b^\dagger - \xi_2 b },
\label{M1_trans}
\ee
where $\xi_2=\xi_1^\ast$ and $\xi_1=  ( 2 \lambda)^{-1} (\thetaone - \I \thetatwo ) $. 
According to Eq.~\eqref{M1_trans}, 
Model 1 is a two-parameter sub-model with a constraint $\xi_2=\xi_1^\ast$ embedded in the four-parameter model. 
We attribute this dependency between $\xi_1$ and $\xi_2$ 
to the complicated bound though it is not clear 
why the change in the bound occurs at $\langle L \rangle_0 = 1 / 2$. 

By adding noise with using the mixed state, the thermal state as the reference state changes 
the feature of the uncertainty relation drastically from the case of the pure state as the reference state.  
In both cases, Model 2 potentially gives more precise way of estimating the position of the electron than Model 1 does. 

\subsubsection{Effects of thermal noise}
We next compare the results between the pure state and the thermal state as the reference state as follows.
First, Model 1 with the pure state as the reference state, the C-R bound has a quasi-classical feature.  
The SLD C-R bound is determined by the constant which is the spread of the position probability density of LLL. 
With the thermal state as the reference state, Model 1 is not D-invariant and 
the uncertainty relation of Model 1 is complicated. The shape of the C-R bound 
depends on the expectation value of angular momentum $\langle L \rangle_0$. 
The SLD  C-R bound becomes achievable only at $\langle L \rangle_0 = 0$. 
Unless this special condition is satisfied, the mixed state model with the thermal noise has discontinuity 
from zero temperature to finite temperature in the quantum C-R bounds. And hence, we cannot simply 
take the zero temperature limit from the thermal state in our model. 

Next, Model 2 with the pure state as the reference state, the generalized RLD exists.  The C-R bound given 
by the generalized RLD is achievable.  The MSE matrix components $V_{1 1}$ and $V_{2 2}$ has a correlation 
as shown in Eq.~\eqref{M2_p_bnd}. 
With the thermal state as the reference state, Model 2 is a simple Gaussian shift model.  Unlike the case of Model 1,
Model 2 with the pure state as the reference state is genuine quantum.
For Model 2, there exists a limit when the temperature goes to zero, or equivalently, 
$\beta\to\infty$ \cite{fujiwara2}. This limit yields the result for the pure-state case studied in Sec.~\ref{sec:Pure_st2}.

\subsubsection{Optical implementation}
The models studied in this paper can naturally be realized in the two-dimensional electron gas at low temperature. 
However, the optimal measurement to attain the quantum C-R bound may not be feasible in such a system. 
Alternatively, one can realize our models in a linear optical system with two modes by tuning parameters properly. 
In this connection, we should not forget to mention related works on parameter estimation problems 
in two mode Gaussian states Refs.~\cite{CI01,Gal13,GL14,BD16,BAL17,BLA18,AKC19}.  
In Ref.~\cite{AKC19}, authors discussed an optimal encoding and measurement scheme 
for estimating two parameters in the pure-state reference state. The optimal state found 
there also comprises of classical correlation of the phase conjugation as in Eq.~\eqref{M1_trans} 
for Model 1. 
The optimal POVM for Model 2 requires measuring 
non-canonical variables. $\pi_x$ and $\pi_y$ \cite{aharonov}.  Therefore, we can only state that Model 2 potentially gives more precise measurement. 
We expect that our result in the thermal state as a reference state should also 
relevant to finding the optimal scheme in the presence of noise. 

\section{Conclusion}\label{sec:5} 
We have investigated the uncertainty relation for estimating $x$ and $y$ components of the position of 
one electron in a uniform magnetic field.  
In the present study, the uncertainty relation upon estimating the expectation values of the two commuting observables, 
$(x, \, y)$ was derived in the framework of the quantum estimation theory.  
As the generators of the unitary transformation, two different sets of generators are used.  
One is a set of canonical momenta, $p_x$ and $p_y$ (Model 1)   
and the other is a set of mechanical momenta, $\pi_x$ and $\pi_y$ (Model 2). 
Based on the analysis by the quantum estimation theory, in both cases, 
we got non-trivial bounds that give the trade-off relations between 
the two commuting observables, $x$ and $y$,  
unlike the result of Heisenberg-Robertson type uncertainty relation.

Although both Model 1 and Model 2 give the same effect to the position probability density defined by the product of 
the wave function and its complex conjugate, the C-R bounds of Model 1 and Model 2 are different for   
pure state and mixed state (thermal state) as the reference state.

With the pure state as the reference state,  
the C-R bound is quasi-classical for Model 1 and 
it is quantum mechanical for Model 2.  
With the thermal state as the reference state, the uncertainty relation given by the C-R bounds is
complicated and the shape of the bounds changes when the expectation value of 
the angular momentum $\langle L \rangle_0$ is 
equal to $1/2$ for Model 1.  Model 2 becomes a simple Gaussian shift model.  
In either case of the pure or thermal state, Model 2 
potentially gives more precise measurement.  

Before closing this paper, we make two remarks.  
First, 
the C-R bound of Model 1 with respect to the thermal state reference state is not achievable except for $\langle L \rangle_0 = 0$.  
A possible extension might be an analysis by minimizing a weighted trace of the mean square error matrix \cite{kull}. 
However, this method gives asymptotically achievable bound only. 
Second, for the thermal state with the $\langle L \rangle_0$ constraint, 
we see the change in the bound shape depending on $\langle L \rangle_0$.  
We have no clue as to why the bound shape changes at $\langle L \rangle_0 = 1/2$ in a simple physical picture so far. 
It should be worthwhile seeing why the bound shape changes there.

\section*{Acknowledgment}
The work is partly supported by JSPS KAKENHI grant number JP17K05571.  
We would like to thank Prof.~Hiroshi Nagaoka for the invaluable discussion and suggestion. 
We would also like to thank anonymous referees for constructive discussions to improve the manuscript. 
%
\appendix

\section{Supplement} \label{sec:suppl}
\subsection{Uncertainty relation by quantum C-R inequality} \label{sec:QCR}
The quantum C-R inequality for the MSE matrix $V_\theta$ is
\be
V_\theta \geq (G_\theta)^{-1}, \label{RLDCR}
\ee
where $G_\theta$ is an arbitrary quantum Fisher information matrix.
Let $(G_\theta)^{-1}$ be
\be
(G_\theta)^{-1} = [{g_\theta}^{i  j}],
\ee 
 
The RLD C-R inequality \eqref{RLDCR} holds iff 
{$\tr \, [V_\theta - (G_\theta)^{-1} ] \geq 0$} and 
$\mathrm{det} \, [V_\theta - (G_\theta)^{-1} ] \geq 0$.  Thus, we have 
\be
V_{ \, 11} - {g_\theta}^{1 1} \geq 0, \quad V_{ \, 22} - {g_\theta}^{22} \geq 0, \nonumber
\ee
and
\be
\mathrm{det} \begin{pmatrix}
V_{11} - {g_\theta}^{1 1} & V_{12} -  {g_\theta}^{1 2}  \\
V_{ 21} - ({g_\theta}^{1 2})^\ast & V_{ 22} - {g_\theta}^{22}  \\
\end{pmatrix} \geq 0, \nonumber
\ee
where $g_\theta^{21}=(g_\theta^{12})^\ast$ is used. The inequality above gives the following inequality.
\be
(V_{ 11} -{g_\theta}^{1 1})(V_{ 22} - {g_\theta}^{22}) \geq | V_{12}  -{g_\theta}^{1 2} \, |^2.  \nonumber
\ee
The right hand side of the inequality above is written as follows.
\begin{align}
 | V_{12}  - {g_\theta}^{1 2} \, |^2 
 &=  | V_{ 12}  - \mathrm{Re} \, {g_\theta}^{1 2} - \I \, \mathrm{Im} \, {g_\theta}^{1 2} \, |^2 \nonumber \\
  &=  | V_{12}  - \mathrm{Re} \, {g_\theta}^{1 2} \, |^2 + | \, \mathrm{Im} \, {g_\theta}^{1 2} \, |^2 \nonumber \\
  & \geq | \, \mathrm{Im} \, {g_\theta}^{1 2}\, |^2.  \nonumber
\end{align}
By applying this inequality, we obtain the following inequalities,
\be
V_{ 11} - {g_\theta}^{1 1} \geq 0, \quad V_{ \, 22} - {g_\theta}^{2 2} \geq 0, \label{V11} 
\ee
\be
(V_{ 11} - {g_\theta}^{1 1})(V_{ 22} - {g_\theta}^{2 2}) \geq | \, \mathrm{Im} \, {g_\theta}^{1 2} \, |^2 . 
\ee
When $\mathrm{Im} \, {g_\theta}^{1 2} = 0$, the uncertainty relation is given by 
Eq.~\eqref{V11} only.
\subsection{Thermal state and Gaussian state} \label{sec:thermal_state}
The thermal state for a single harmonic oscillator, $\rho_\beta$ is described as 
\be
\rho_\beta = Z_\beta^{-1} \e^{-\beta H},
\ee
where $Z_\beta = \tr \, [\e^{-\beta H}]$ and $\beta = T^{-1}$.  $T$ is temperature. 

By using Hamiltonian $H=\omega \, (a^\dagger a + 1 / 2)$ and $a^\dagger a \ket{n} = n \ket{n}$, 
$\, \e^{- \beta H}$ is
\be
\e^{- \beta H} 
=  \sum_{n=0}^\infty \e^{-\beta H} \ket{n} \bra{n} \nonumber \\
=  \e^{- \frac{1}{2} \beta \omega } \sum_{n=0}^\infty  \gamma^n \ket{n} \bra{n}, \nonumber 
\ee
where $\gamma = \mathrm{e}^{-\beta \omega} $. $Z_\beta$ is
\be
Z_\beta 
= \tr \, [\e^{-\beta H}]  \nonumber \\
= \frac{\e^{-  \frac{1}{2} \beta \omega }}{1-\gamma} . \nonumber 
\ee
We obtain
\be
\rho_\beta = Z_\beta^{-1} \e^{-\beta H} = (1 - \gamma) \sum_n \gamma^n \ket{n} \bra{n}. \nonumber
\ee
We first calculate the matrix element of $\rho_\beta$ by the basis as the Gaussian state, $\braket{z_1 | \rho_\beta | z_2}$.  Next, we make the same matrix element of the Gaussian state to see if they match.

$\braket{z_1 | \rho_\beta | z_2}$ is
\begin{align}
\braket{z_1 | \rho_\beta | z_2} 
&= (1-\gamma) \sum_n \gamma^n \braket{z_1 | n} \braket{n | z_2 }  \nonumber \\
&= (1-\gamma) \mathrm{e}^{-\frac{1}{2} | z_1 |^2 -\frac{1}{2} | z_2 |^2 + \gamma z_1^\ast z_2}.
\label{rho_mat}
\end{align}
The Gaussian state $S_\kappa$ is defined by
\be
S_\kappa = \frac{1}{2 \pi \kappa^2} \int \e^{- \frac{| z |^2}{2 \kappa^2}} \ket{ z} \bra{ z} d^2  z.
\nonumber
\ee
Then its matrix element $ \bra{z_1} S_\kappa \ket{z_2}$ is
\begin{align} 
 \bra{z_1} S_\kappa \ket{z_2}
&= \frac{1}{2 \pi \kappa^2} \int \e^{- \frac{| z |^2}{2 \kappa^2}} \braket{z_1 | z}  
\braket{ z | z_2} d^2 z \nonumber \\
&=\frac{1}{2 \pi \kappa^2} \int \e^{-(\frac{1}{2\kappa^2} +1)  |  z |^2 + z_1^\ast z_2 + 
z_1 z_2^\ast} d^2 z \, \e^{-\frac{1}{2} | z_1 |^2 -\frac{1}{2} | z_2 |^2 }.  \nonumber 
\end{align}
By using
\be
\int \e^{- \alpha | z |^2 + \beta z + \gamma z^\ast} d^2 z
= \frac{\pi}{\alpha} \e^{\frac{\beta \gamma}{\alpha}}, \nonumber
\ee
we obtain
\be
\bra{z_1} \rho_\beta \ket{z_2}
= \frac{1}{2 \kappa^2+1} \e^{-\frac{1}{2} | z_1 |^2 -\frac{1}{2} | z_2 |^2 + \frac{z_1^\ast z_2}{\frac{1}{2 \kappa^2}+1}} .
\label{Sk_mat}
\ee
From \eqref{rho_mat} and \eqref{Sk_mat}, $\braket{z_1 | S_\kappa | z_2} = \bra{z_1}  \rho_\beta \ket{z_2}$ holds
iff
\be
2 \kappa^2 = \dfrac{\gamma}{1-\gamma} = \frac{1}{\e^{\beta \omega} - 1}.
\label{kappa}
\ee
Therefore, we obtain
\be
\rho_\beta = \frac{1}{2 \pi \kappa^2} \int \e^{- \frac{| z^\prime |^2}{2 \kappa^2}} \ket{ z} \bra{ z} d^2  z.
\label{thermal}
\ee
where $2\kappa^2$ is given by Eq.~\eqref{kappa}.
\subsection{Model 1 unitary transformation \\ 
in terms of the creation and annihilation operators} \label{sec:alt_rep}
The unitary transformations of the Model 1, $ \e^{- \I \thetaone p_x}  \e^{- \I \thetatwo p_y}$ is
\be
 \e^{- \I \thetaone p_x} \e^{- \I \thetatwo p_y}
= \e^{\frac{1}{2 \lambda} \, \{ (a^\dagger- a) + (b^\dagger - b) \} \thetaone}
\e^{- \frac{\I}{2 \lambda} \, \{ (a^\dagger + a) - (b^\dagger + b) \} \thetatwo}. \nonumber 
\ee
Since $p_x$ and $p_y$ commute, 
\be
 \e^{- \I \thetaone p_x} \e^{- \I \thetatwo p_y}
= \e^{- \I  \thetaone p_x - \I \thetatwo p_y} 
= \e^{\frac{1}{2 \lambda} \, \{ (a^\dagger- a) + (b^\dagger - b) \} \thetaone \} 
- \frac{\I}{2 \lambda} \, \{ (a^\dagger + a) - (b^\dagger + b) \} \thetatwo }. \nonumber 
\ee
Therefore, 
\be
 \e^{- \I \thetaone p_x} \e^{- \I \thetatwo p_y}
=  \mathrm{e}^{ \xi a^\dagger - \xi^\ast a }  \mathrm{e}^{ \xi^\ast b^\dagger - \xi b }, \label{unitary_p}
\ee
where $\xi = (2 \lambda)^{-1} (\thetaone - \I \thetatwo )$.

\section{Calculation} \label{sec:calc}
\subsection{SLD and RLD: The thermal state as the reference state} \label{sec:SLD_RLD}
First, we briefly explain that SLD and RLD Fisher information matrices for the mixed state are 
independent of the parameters 
$\theta = (\thetaone, \, \thetatwo)$ in the unitary transformation $U(\thetaone, \, \thetatwo)$.

Let Model 1 SLD and Model 1 RLD of Model 1 be $L^{(1)}_{\mathrm{S}, \, j}(\theta)$ and $L^{(1)}_{R, \, j}(\theta)$, respectively.
With using the unitary transformation  $U(\thetaone, \, \thetatwo) = \e^{- \I \thetaone p_x} \e^{- \I \thetatwo p_y}$,  
$L^{(1)}_{\mathrm{S}, \, j}(0)$ and $L^{(1)}_{\mathrm{R}, \, j}(0)$ are written as
\begin{align}
L^{(1)}_{\mathrm{S}, \, j}(\theta) &= U(\thetaone, \, \thetatwo) \, L^{(1)}_{\mathrm{S}, \, j}(0) \, U^\dagger (\thetaone, \, \thetatwo), 
\nonumber \\
L^{(1)}_{\mathrm{R}, \, j}(\theta) &= U(\thetaone, \, \thetatwo) \, L^{(1)}_{\mathrm{R}, \, j}(0) \, U^\dagger (\thetaone, \, \thetatwo). 
\nonumber
\end{align}
For the RLD Fisher information $G^{(1)}_\mathrm{R} (\theta) = [g^{(1)}_{\mathrm{R}, \: i j}(\theta)]$, we can derive the relation below 
if the transformation is unitary.
\be
g^{(1)}_{\mathrm{R}, \: i j}(\theta) =\tr \, [ \rho_\theta \, L^{(1)}_{\mathrm{R}, \, j}(\theta) L^{(1)\, \dagger}_{\mathrm{R}, \, i}(\theta)]
= \tr \, [ \rho_0 \, L_{\mathrm{R}, \, j}(0) L^{(1) \, \dagger}_{\mathrm{R}, \, i}(0) ]. \nonumber
\ee
If the transformation is unitary, the RLD Fisher information $G_\mathrm{R} (\theta)$ does not depend on the parameters 
$\thetaone$ and $\thetatwo$.  
Then, we can write $G_\mathrm{R} (\theta)$ as $G_\mathrm{R} = [ g_{\mathrm{R}, \: i j} ]$.  
From Eqs.~(\ref{RLD_Fisher}, ~\ref{SLD_Fisher}), we can show that the same holds for  the SLD Fisher information $G_\mathrm{S} (\theta)$. 
Therefore, if we have $L^{(1)}_{\mathrm{S}, \, i}(0)$ and $L^{(1)}_{\mathrm{R}, \, i}(0)$, it is enough to obtain the SLD and RLD Fisher information matrices.  The same is true for Model 2. 

\subsubsection{Model 1 SLD: $L^{(1)}_{S, \, 1}(0), \, L^{(1)}_{S, \, 2}(0), Z \: matrix \: Z^{p \, \mathrm{thermal}}$}  
\begin{align}
L^{(1)}_{\mathrm{S}, \, 1}(0) &= \frac{1}{\lambda (1 + 4 \ka^2)} (a + a^\dagger) 
+ \frac{1}{\lambda (1 + 4 \kb^2)} (b + b^\dagger), \nonumber \\
L^{(1)}_{\mathrm{S}, \, 2}(0) &= \frac{ \I}{\lambda (1 + 4 \ka^2)} (a - a^\dagger) 
- \frac{ \I}{\lambda (1 + 4 \kb^2)} (b - b^\dagger). \nonumber
\end{align}
With using $p_x, \, p_y$ and $x, \, y$, 
\begin{align}
L^{(1)}_{\mathrm{S}, \, 1}(0) &= ( \frac{1}{ 1 + 4 \ka^2} + \frac{1}{ 1 + 4 \kb^2} ) p_y
+ \frac{1}{\lambda^2}( \frac{1}{ 1 + 4 \ka^2} - \frac{1}{ 1 + 4 \kb^2} ) \, x, \nonumber \\
L^{(1)}_{\mathrm{S}, \, 2}(0) &= - ( \frac{1}{ 1 + 4 \ka^2} - \frac{1}{ 1 + 4 \kb^2} ) p_x
+ \frac{1}{\lambda^2}( \frac{1}{ 1 + 4 \ka^2} + \frac{1}{ 1 + 4 \kb^2} ) \, y. \nonumber
\end{align}
The SLD Fisher information matrix $\Gspt$ is calculated as
\be
\Gspt
= \frac{1}{\lambda^2} ( \frac{1}{1 + 4 \kappa_a^2} + \frac{1}{1 + 4 \kappa_b^2})
\begin{pmatrix}
1 &  0 \\
0  & 1 \\
\end{pmatrix}. \nonumber
\ee
Z matrix $Z^{p \, \mathrm{thermal}}$ is calculated as
\be
Z^{p \, \mathrm{thermal}}
=  \frac{\lambda^2}{ 1 + 2 \ka^2 + 2 \kb^2} 
\begin{pmatrix}
  \frac{1}{2} + 2 \ka^2 + 2 \kb^2 + 8 \ka^2 \kb^2 & \I   \, (2 \kb^2 - 2 \ka^2) \\
- \I   \, (2 \kb^2 - 2 \ka^2) &     \frac{1}{2} + 2 \ka^2 + 2 \kb^2 + 8 \ka^2 \kb^2  \\
\end{pmatrix}. \nonumber
\ee
From this expression, we have
\be
Z^{p \, \mathrm{thermal}} = (\Grpt)^{-1} +  \Delta g
\begin{pmatrix}
  1 & 0 \\
  0 & 1  \\
\end{pmatrix}. \nonumber
\ee
Since $\Delta g\neq 0$, we see that $Z^{p \, \mathrm{thermal}}\neq (\Grpt)^{-1}$. 
This implies the model is not D-invariant \cite{suzuki}. 
\subsubsection{Model 2 SLD: $L^{(2)}_{\mathrm{S}, \, 1}(0), \, L^{(2)}_{\mathrm{S}, \, 2}(0), 
Z \: matrix \: Z^{\pi \, \mathrm{thermal}}$}
\begin{align}
L^{(2)}_{\mathrm{S}, \, 1}(0) &= \frac{2}{\lambda (1 + 4 \ka^2)} (a + a^\dagger), \nonumber \\
L^{(2)}_{\mathrm{S}, \, 2}(0) &= \frac{2 \I}{\lambda (1 + 4 \ka^2)} (a - a^\dagger). \nonumber
\end{align}
With using $p_x, \, p_y$ and $x, \, y$, 
\begin{align}
L^{(2)}_{\mathrm{S}, \, 1}(0) &= \frac{2}{1 + 4 \ka^2} ( p_y + \frac{1}{\lambda^2} x), \nonumber \\
L^{(2)}_{\mathrm{S}, \, 2}(0) &= \frac{2}{1 + 4 \ka^2} ( - p_x + \frac{1}{\lambda^2} y). \nonumber
\end{align} 
The inverse of the SLD Fisher information matrix $\Gspit $ is 
\be
\Gspit
= \frac{4}{\lambda^2 (1 + 4 \ka^2)} \begin{pmatrix}
1   & 0 \\
0 & 1   \\
\end{pmatrix}. \nonumber
\ee
Z matrix $Z^{\pi \, \mathrm{thermal}}$ is 
\be
Z^{\pi \, \mathrm{thermal}}
= \frac{\lambda^2}{4} \begin{pmatrix}
1 + 4 \ka^2  & \I \\
- \I & 1 + 4 \ka^2  \\
\end{pmatrix}. \nonumber
\ee
\subsubsection{Model 1 RLD: $L^{(1)}_{\mathrm{R}, \, 1}(0), \, L^{(1)}_{\mathrm{R}, \, 2}(0)$}  
\begin{align}
L^{(1)}_{\mathrm{R}, \, 1}(0) &= \frac{1}{2\lambda} ( \frac{1}{1 + 2 \ka^2} a + \frac{1}{2 \ka^2} a^\dagger) 
+ \frac{1}{2\lambda} ( \frac{1}{1 + 2 \kb^2} b + \frac{1}{2 \kb^2} b^\dagger) , \nonumber \\
L^{(1)}_{\mathrm{R}, \, 2}(0) &= - \frac{\I}{2\lambda} ( - \frac{1}{1 + 2 \ka^2} a + \frac{1}{2 \ka^2} a^\dagger) 
+ \frac{\I}{2\lambda} ( - \frac{1}{1 + 2 \kb^2} b + \frac{1}{2 \kb^2} b^\dagger)  . \nonumber
\end{align}
The RLD Fisher information matrix $\Grpt$ is calculated as
\be
\Grpt =  \frac{1}{4 \lambda^2} 
\begin{pmatrix}
 \frac{1}{ 1 + 2 \kappa_a^2} + \frac{1}{ 2 \kappa_a^2} + \frac{1}{ 1 + 2 \kappa_b^2} + \frac{1}{ 2 \kappa_b^2} & - \I   \, [ \frac{1}{ 2 \kappa_a^2 (1 + 2 \kappa_a^2)} -
 \frac{1}{ 2 \kappa_b^2 (1 + 2 \kappa_b^2)} ] \\
 \I  \, [ \frac{1}{ 2 \kappa_a^2 (1 + 2 \kappa_a^2)} 
- \frac{1}{ 2 \kappa_b^2 (1 + 2 \kappa_b^2)}  ]  &     \frac{1}{ 1 + 2 \kappa_a^2} + \frac{1}{ 2 \kappa_a^2} + \frac{1}{ 1 + 2 \kappa_b^2} + \frac{1}{ 2 \kappa_b^2}  \nonumber 
\end{pmatrix}.
\ee
\subsubsection{Model 2 RLD: $L^{(2)}_{\mathrm{R}, \, 1}(0), \, L^{(2)}_{\mathrm{R}, \, 2}(0)$}
\begin{align}
L^{(2)}_{\mathrm{R}, \, 1}(0) &= \frac{1}{\lambda} ( \frac{1}{1 + 2 \ka^2} a + \frac{1}{2 \ka^2}  a^\dagger), 
\nonumber \\
L^{(2)}_{\mathrm{R}, \, 2}(0) &= \frac{\I}{\lambda} ( \frac{1}{1 + 2 \ka^2} a - \frac{1}{2 \ka^2}  a^\dagger). 
\nonumber
\end{align}
The creation annihilation operators, $a, \, a^\dagger$ and $b, \, b^\dagger$ are written as follows.
\begin{align}
a &= \frac{\lambda}{2} [ ( \I p_x + p_y) + \frac{1}{\lambda^2} (x - \I y)], \nonumber \\
a^\dagger &= \frac{\lambda}{2} [ ( - \I p_x + p_y) + \frac{1}{\lambda^2} (x + \I y)], \nonumber \\
b &= \frac{\lambda}{2} [ (  \I p_x - p_y) + \frac{1}{\lambda^2} (x + \I y)], \nonumber \\
b^\dagger &= \frac{\lambda}{2} [ ( - \I p_x - p_y) + \frac{1}{\lambda^2} (x - \I y)]. \nonumber 
\end{align}
The RLD Fisher information matrix $\Grpit $ is 
\be
\Grpit 
= \frac{1}{\lambda^2} \frac{1}{2 \ka^2 (1+ 2\ka^2)} \begin{pmatrix}
1 + 4 \ka^2  & - \I \\
 \I & 1 + 4 \ka^2  \\
\end{pmatrix} . \nonumber
\ee

\subsection{SLD, Generalized RLD: Pure state as the reference state} \label{sec:gen_RLD}
Let the SLD of a pure state $\rho_\theta = \ket{\psi_\theta} \bra{\psi_\theta}$ be $L_{S, \, i}$.  
Then,  $L_{\mathrm{S}, \, i}$ is expressed as~\cite{fujiwara2},  
\be
L_{\mathrm{S}, \, i} = 2 {\partial_i } \rho_\theta = 2 \partial_i ( \ket{\psi_\theta} \bra{\psi_\theta}). \label{p_SLD}
\ee
\subsubsection{Model 1 SLD: $L^{(1)}_{\mathrm{S}, \, 1}(\theta), \: L^{(1)}_{\mathrm{S}, \, 2}(\theta)$}\label{sec:gen_RLD1}
We set the reference state $\rho_0$ as $\rho_0 = \ket{0, \, 0} \bra{0, \, 0}$.  
From Eq.~\eqref{c_ope22}, $\rho^{p}_\theta$ is expressed as
\begin{align}
\rho^{p}_\theta 
&= \e^{- \I \thetaone p_x} \e^{- \I \thetatwo p_y} \ket{0, \, 0} \bra{0, \, 0} \e^{ \I \thetatwo p_y} \e^{ \I \thetaone p_x} \\
&= U(\theta) \ket{0, \, 0} \bra{0, \, 0} U^\dagger(\theta).
\end{align}
where $U(\theta) =\e^{- \I \thetaone p_x} \e^{- \I \thetatwo p_y} $.  From Eq.~\eqref{p_SLD}, 
the SLD's of Model 1 are expressed as 
\begin{align}
L^{(1)}_{\mathrm{S}, \, 1}(0) &=  - 2 \I [ p_x, \, \rho_0] \nonumber, \\
L^{(1)}_{\mathrm{S}, \, 2}(0) &=  - 2 \I [ p_y, \, \rho_0] \nonumber. 
\end{align}
where $L^{(1)}_{\mathrm{S}, \, j}(\theta) = U(\theta) L^{(1)}_{\mathrm{S}, \, j}(0) U^\dagger (\theta), \: (j = 1, \,2)$. 

By using Eq.~\eqref{p}, $L^{(1)}_{S, \, 1}(0)$ and $L^{(1)}_{\mathrm{S}, \, 2}(0)$ are also written as 
\begin{align}
L^{(1)}_{\mathrm{S}, \, 1}(0) &=  \frac{1}{\lambda} [ ( a^\dagger - a ) + ( b^\dagger - b ) , \, \rho_0 ] \nonumber, \\
L^{(1)}_{\mathrm{S}, \, 2}(0) &=  - \frac{\I}{\lambda} [ ( a^\dagger + a ) - ( b^\dagger + b ), \, \rho_0] \nonumber. 
\end{align}
With these SLD's, the Fisher information matrix $\Gspp$ is calculated as 
\be
\Gspp  = \frac{2}{\lambda^2} \begin{pmatrix}
1  & 0 \\
0 & 1  \\
\end{pmatrix}.   \nonumber
\ee
From the direct calculation of Eq.~\eqref{geneRLD}, we can show that $\Gspp = \GGrpp$. 
\subsubsection{Model 2 SLD: $L^{(2)}_{\mathrm{S}, \, 1}(\theta), \: L^{(2)}_{\mathrm{S}, \, 2}(\theta)$}\label{sec:gen_RLD2}
From Eq.~\eqref{c_ope22}, $\rho^{\pi}_\theta$ is expressed as
\be
\rho^{\pi}_\theta 
= \e^{- \I \thetaone \pi_x} \e^{- \I \thetatwo \pi_y} \ket{0, \, 0} \bra{0, \, 0} \e^{ \I \thetatwo \pi_y} \e^{ \I \thetaone \pi_x}.
\label{state_2}
\ee
The unitary transformation $\e^{- \I \thetaone \pi_x} \e^{- \I \thetatwo \pi_y}$ is calculated as follows~\cite{klauder}.
\be
\e^{- \I \thetaone \pi_x} \e^{- \I \thetatwo \pi_y} = \e^{\frac{1}{\lambda^2} \thetaone \thetatwo} 
\e^{- \I \thetaone \pi_x  - \I \thetatwo \pi_y  } .
\label{u_tr1}
\ee
By substituting Eq.~\eqref{u_tr1} in Eq~\eqref{state_2}, we obtain
\be
\rho^{\pi}_\theta 
= U(\theta) \ket{0, \, 0} \bra{0, \, 0} U^\dagger(\theta).
\label{state_3}
\ee
where $U(\theta) =\e^{- \I \thetaone \pi_x  - \I \thetatwo \pi_y  }$.

From Eq.~\eqref{p_SLD}, 
the SLD's of Model 2 are expressed as 
\begin{align}
L^{(2)}_{\mathrm{S}, \, 1}(0) &=  -2 \I [ \pi_x, \, \rho_0] \nonumber, \\
L^{(2)}_{\mathrm{S}, \, 2}(0) &=  -2 \I [ \pi_y, \, \rho_0] \nonumber. 
\end{align}
where $L^{(2)}_{\mathrm{S}, \, j}(\theta) =  U(\theta) L^{(2)}_{S, \, j}(0) U^\dagger (\theta), \: (j = 1, \,2)$. 

By using Eq.~\eqref{pi}, 
\begin{align}
L^{(2)}_{\mathrm{S}, \, 1}(0) &=  - \frac{2 \I}{\lambda} [ a^\dagger + a   , \, \rho_0] \nonumber, \\
L^{(2)}_{\mathrm{S}, \, 2}(0) &=   \frac{2}{\lambda} [ a^\dagger - a , \, \rho_0] \nonumber. 
\end{align}
Thus, the SLD Fisher information $\Gspip$ is
\be
\Gspip = \frac{4}{\lambda^2} \begin{pmatrix}
1  & 0 \\
0 & 1  \\
\end{pmatrix}. \nonumber
\ee 
From Eq.~\eqref{p_SLD}, the generalized RLD Fisher information $\GGrpip$ is 
\be
\GGrpip = \frac{4}{\lambda^2} \begin{pmatrix}
1  & \I \\
-\I & 1  \\
\end{pmatrix}. \nonumber
\ee
%


\end{document}